\documentclass[prd,amsmath,amssymb,superscriptaddress,preprintnumbers,twocolumn,10pt]{revtex4-1}%superscriptaddress,showpacs,

\usepackage{graphicx}
\usepackage{dcolumn}
\usepackage{bm}
\usepackage{amssymb}
\usepackage{latexsym}
\usepackage{booktabs}
\usepackage{amsmath}
\usepackage{multirow}
\usepackage{url}
\usepackage{footnote}
\usepackage{float}
\usepackage[colorlinks=true, linkcolor=red, citecolor=blue]{hyperref}

\usepackage[normalem]{ulem}
\usepackage{color}
\usepackage{array}
\usepackage{enumerate}

\begin{document}

\title{Robust inference of gravitational wave source parameters in the presence of noise transients using normalizing flows}

\author{Chun-Yu Xiong}
%\affiliation{Key Laboratory of Cosmology and Astrophysics (Liaoning) \& College of Sciences, Northeastern University, Shenyang 110819, China}
\affiliation{Liaoning Key Laboratory of Cosmology and Astrophysics, \ College of Sciences, Northeastern University, Shenyang 110819, China}
\author{Tian-Yang Sun}
\affiliation{Liaoning Key Laboratory of Cosmology and Astrophysics, \ College of Sciences, Northeastern University, Shenyang 110819, China}

\author{Jing-Fei Zhang}
\affiliation{Liaoning Key Laboratory of Cosmology and Astrophysics, 
\ College of Sciences, Northeastern University, Shenyang 110819, China}
\author{Xin Zhang}\thanks{Corresponding author}\email{zhangxin@mail.neu.edu.cn}
%\author{Xin Zhang\footnote{Corresponding author}}
%\email{zhangxin@mail.neu.edu.cn}
\affiliation{Liaoning Key Laboratory of Cosmology and Astrophysics, \ College of Sciences, Northeastern University, Shenyang 110819, China}

\affiliation{MOE Key Laboratory of Data Analytics and Optimization for Smart Industry, \ Northeastern University, Shenyang 110819, China}

\affiliation{National Frontiers Science Center for Industrial Intelligence and Systems Optimization, \ Northeastern University, Shenyang 110819, China}

%\affiliation{Key Laboratory of Data Analytics and Optimization
%for Smart Industry, (Northeastern University), Ministry of Education, Shenyang 110819, China}
%\affiliation{Center for High Energy Physics, Peking University, Beijing 100080, China}
%\date{\today}
%\cite{LIGOScientific:2016gtq, Nuttall:2018xhi, Cabero:2019orq, LIGO:2021ppb, LIGOScientific:2021tfm}
 %\cite{LIGOScientific:2016gtq, Nuttall:2018xhi, Cabero:2019orq, LIGO:2021ppb, LIGOScientific:2021tfm}, current methods cannot completely remove it from the GW signal.

\begin{abstract}
Gravitational wave (GW) detection is of paramount importance in fundamental physics and GW astronomy, yet it presents formidable challenges. One significant challenge is the removal of noise transient artifacts known as ``glitches'', which greatly impact the search and identification of GWs. Recent research has achieved remarkable results in data denoising, often using effective modeling methods to remove glitches. However, for glitches from uncertain or unknown sources, current methods cannot completely eliminate them from the GW signal. In this work, we leverage the inherent robustness of machine learning to obtain reliable posterior parameter distributions directly from GW data contaminated by glitches. Our network model provides reasonable and rapid parameter inference even in the presence of glitches, without needing to remove them. We also investigate various factors affecting the rationality of parameter inference in our normalizing flow network, including glitch and GW parameters. The results demonstrate that the normalizing flow can reasonably infer the source parameters of GWs even with unknown contamination. We find that the nature of the glitch itself is the only factor that can affect the rationality of the inferred results. With improvements to our model, we anticipate accelerating the localization of electromagnetic counterparts and providing priors for more accurate deglitching, thereby speeding up subsequent data processing procedures.

%Gravitational wave (GW) detection is of paramount importance in fundamental physics and GW astronomy, but it also presents formidable challenges. One of the challenging aspects is the removal of noise transient artifacts called ``glitch", which significantly impacts the search and identification of GW. While recent research has achieved remarkable results in data denoising, they often employ effective modeling methods to remove glitches. For uncertain or unknown sources of glitches, current methods cannot completely remove them from the GW signal. In this work, we use the inherent robustness of machine learning to obtain reliable posterior parameter distributions directly from GW data contaminated by glitches. In the presence of glitches contamination, our network model provides reasonable and fast parameter inference without removing the glitches. Additionally, we investigated various factors that can affect the rationality of parameter inference in our normalizing flow network, including glitch parameters and GW parameters. The results show that the normalizing flow can infer the source parameters of GWs reasonably even in the case of unknown contamination. We find that only the nature of the glitch itself can affect the rationality of the inferred results. With the improvement of our model, we anticipate speeding up the localization of electromagnetic counterparts and providing priors for more accurate deglitching, thereby accelerating subsequent data processing procedures.

\end{abstract}
%\pacs{95.36.+x, 98.80.Es, 98.80.-k}

\maketitle
\section{Introduction}

%Approximately a hundred years ago, Einstein predicted the existence of gravitational waves (GWs) in the theory of general relativity. Until 2015, the Advanced LIGO \cite{VIRGO:2012oxz} and Virgo \cite{LIGOScientific:2013pcc} GW detectors have made the first direct detection of GW signal from the merger of a binary black hole (BBH) system \cite{LIGOScientific:2017vox, LIGOScientific:2017ycc}, GW150914 \cite{LIGOScientific:2016aoc}. GW research plays a crucial role in testing general relativity \cite{Berti:2015itd,LIGOScientific:2016lio,LIGOScientific:2018dkp,LIGOScientific:2019fpa,LIGOScientific:2020tif,LIGOScientific:2021sio,Gong:2021jgg,Gong:2023ffb}, probing cosmological phenomena, and accurately measuring cosmological parameters \cite{LIGOScientific:2017adf,Chen:2017rfc,LIGOScientific:2021aug,DES:2019ccw,DES:2020nay,Sakstein:2017xjx, Verde:2019ivm, Palmese:2021mjm, Jin:2020hmc, Wang:2021srv, Jin:2021pcv, Song:2022siz, Li:2023gtu, Jin:2023sfc,LIGOScientific:2019zcs}, contributing to our understanding of fundamental physics and the evolution of the universe.

%Approximately a hundred years ago, Einstein predicted the existence of gravitational waves (GWs) within the framework of general relativity. It wasn't until 2015 that the Advanced LIGO and Virgo GW detectors achieved the first direct detection of a GW signal from the merger of a binary black hole (BBH) system, known as GW150914 \cite{LIGOScientific:2016aoc}. This discovery marked a monumental advance in the field.
Approximately a hundred years ago, Einstein predicted the existence of gravitational waves (GWs) within the framework of general relativity. {It wasn't until 2015 that the Advanced LIGO detectors achieved the first direct detection of a GW signal from the merger of a binary black hole (BBH) system, known as GW150914 \cite{LIGOScientific:2016aoc}.} This discovery marked a monumental advance in the field.

GW research is pivotal for testing general relativity \cite{Berti:2015itd,LIGOScientific:2016lio,LIGOScientific:2018dkp,LIGOScientific:2019fpa,LIGOScientific:2020tif,LIGOScientific:2021sio,Gong:2021jgg,Gong:2023ffb}, exploring astrophysical and cosmological phenomena, and accurately measuring cosmological parameters \cite{LIGOScientific:2017adf,Sakstein:2017xjx,Chen:2017rfc,DES:2019ccw,Verde:2019ivm,Jin:2020hmc,DES:2020nay,LIGOScientific:2019zcs,Jin:2021pcv,Wang:2021srv,guo2022standard,Jin:2022tdf,LIGOScientific:2021aug,Palmese:2021mjm,Song:2022siz,Li:2023gtu,Jin:2023sfc,Han:2023exn,Jin:2023tou}. These activities significantly enhance our understanding of fundamental physics and the evolution of the universe.

However, the data received by detectors is highly susceptible to interference from noise transient, resulting in deviation in the analysis process \cite{ LIGOScientific:2017tza,LIGOScientific:2018mvr, LIGOScientific:2020ibl, KAGRA:2021vkt}. The LIGO-Virgo-KAGRA collaboration has revealed that non-Gaussian features in the data, such as noise transient, or called ``glitch'' \cite{LIGOScientific:2019hgc, LIGO:2021ppb, Virgo:2022ysc}, can significantly impact the detector's ability to search for GW signals \cite{LIGOScientific:2019hgc}.
 %Indeed, interference, including glitches, can have various effects on GW data, which can impact subsequent data analysis in significant ways. 
%They are categorized according to time, frequency, signal-to-noise ratio (SNR), and duration \cite{Robinet:2020lbf,Lynch:2015yin}, with the most common being blip glitches, scattered-light glitches and thunderstorm glitches

Current observations have characterized dozens of different glitches by Gravity Spy \cite{Zevin:2016qwy, Soni:2021cjy}. These glitches can exhibit variations in terms of their frequency, amplitude, and duration, making their identification and mitigation complex tasks. {They are categorized according to time, frequency, signal-to-noise ratio (SNR), and duration \cite{Robinet:2020lbf,Lynch:2015yin}, and there are many kinds of glitches, among which the common glitches being blip glitches, scattered-light glitches and thunderstorm glitches \cite{Razzano:2018fxb,  Hourihane:2022doe,Zevin:2023rmt}.} Indeed, glitches have significant impacts on GW data and subsequent data analysis \cite{LIGOScientific:2014oec,LIGOScientific:2016vlm,LIGOScientific:2016dsl}. These non-Gaussian non-stationary features can mimic or mask the signature of GWs, making it more difficult to distinguish real signals from noise \cite{Nitz:2017lco, Cabero:2019orq}. For example, they can affect the search phase, causing detectors to generate false alarms \cite{LIGOScientific:2017tza}. The presence of glitches causes incorrect sky positioning in GW data analysis, affects the tracking of electromagnetic (EM) counterparts \cite{Pankow:2018qpo}, and introduces biases in the inference of cosmological parameters \cite{Macas:2022afm, Zhu:2023jti}. 

The interference caused by glitches in GW data cannot be ignored, and many efforts are being made to mitigate their disturbance \cite{Powell:2018csz,LIGO:2020zwl,Cornish:2021wxy,Talbot:2021igi,Steltner:2021qjy,Udall:2022vkv,Hourihane:2022doe,Mohanty:2023mjn}. The gating approach uses a windowing function to directly smooth the portion affected by the glitch for events whose time-frequency can be separated \cite{Usman:2015kfa}. {\tt Gwsubtract} \cite{Davis:2018yrz,KAGRA:2021vkt,Davis:2022ird} relies on the auxiliary channel to remove glitches, but it has a limited impact on removing glitches that do not appear in auxiliary channels. {\tt BayesWave} reconstructs the Gaussian noise, glitches, and GW signals using wavelet transforms and inter-correlations between detectors to remove glitches \cite{Cornish:2014kda,Littenberg:2015kpb,Cornish:2020dwh}. However, this high-dimensional Bayesian analysis is extremely computationally intensive, and individual events cannot be analyzed in a timely and efficient manner at expected future event rates. Therefore, the development of an effective framework for rapidly mitigating the impacts of glitches and inferring GW parameters has become an urgent need.

%For events where time-frequency can be separated, gating can be used for fast analysis, as in GW170817. In situations where the temporal and frequency overlap becomes non-negligible, a technique known as ``global fitting" is commonly employed. One prominent example of such a method is the utilization of BayesWave. BayesWave enables the simultaneous simulation and analysis of Gaussian noise, non-Gaussian noise artifacts, and short-duration GW transients, providing a comprehensive approach to extracting valuable information from the data.

The development of Artificial Intelligence offers an excellent way to facilitate GW research by overcoming the problems of high computational costs and handling huge amounts of data \cite{Huerta:2020xyq}. Currently, most state-of-the-art machine learning algorithms remove non-Gaussian noise components \cite{Biswas:2013wfa,Vajente:2019ycy,Ormiston:2020ele,Merritt:2021xwh,Yu:2021swq,Ashton:2022ztk,Bini:2023gil} by fitting their parameters. Moreover, machine learning techniques \cite{Gabbard:2017lja,Chua:2018woh,Wang:2019zaj,Chen:2020ehw,Cabero:2020eik,Green:2020dnx,Gunny:2021gne,Ruan:2021fxq,Zhao:2022qob,Dax:2022pxd,Ruan:2023fce,Yun:2023vwa,Ma:2023evh,Zhao:2023ncy} utilize the normalizing flow \cite{Williams:2021qyt,Ruhe:2022ddi,Langendorff:2022fzq,Du:2023plr,Sun:2024ywb} for the posterior distribution of source parameters \cite{Fan:2018vgw,Chua:2019wwt,Dax:2021tsq,Gabbard:2019rde}. While these methods can effectively enhance data quality \cite{Zhao:2024btd} in certain aspects, they cannot ensure that the noise remaining after deglitching consists solely of Gaussian noise. Moreover, other approaches have not involved the removal or subtraction of non-Gaussian noise \cite{Du:2023plr,Sun:2023prd}.
%Instead, these approaches focus on incorporating non-Gaussian noise into the training process and utilizing likelihood-free methods to directly infer GW parameters. 
%Recently, there have been some efforts to bypass the removal of non-Gaussian noise and instead utilize machine learning methods to directly infer the probability distribution of parameter values \cite{Du:2023plr,Sun:2023prd}. \cite{Gabbard:2019rde,Dax:2021tsq}

In our previous work, we employed the normalizing flow to perform parameter inference on GW signals contaminated by several common glitches \cite{Sun:2023vlq}. However, glitches come in different types with varying frequencies of occurrence, and not all types of glitches can be well modeled or have sufficient numbers for training. We are seeking a method that does not rely on modeling to remove glitches and enable rapid parameter inference. Recent studies have demonstrated the robustness of machine learning \cite{Wei:2019zlc,Chatterjee:2021lit,Wang:2022quo,Wang:2023lif,Xu:2024jbo}. The robustness of machine learning methods can aid in performing parameter inference reasonably, even without explicitly modeling and removing glitches. It enables the effective mitigation of non-Gaussian noise interference while inferring GW parameters. %We can use Gaussian noise for training to infer signal parameters contaminated by non-Gaussian noise. It is foreseeable that this approach can infer the source parameter of GWs contaminated by glitches originating from unknown sources.

In this work, our primary aim is to investigate the impact of noise transients on the parameter inference obtained from the normalizing flow network. We use Gaussian noise for training to infer signal parameters contaminated by non-Gaussian noise. Specifically, we utilize simulated sine-Gaussian glitches, encompassing a wide range of morphologies found in advanced detector noise transients \cite{Powell:2015ona, Bose:2016jeo}. We discuss the effect of the number of detectors on robust inference in the normalizing flow network. We find that our model provides a relatively better posterior distribution compared to the Bayesian inference method when dealing with glitch-contaminated GW data. In this work, we utilize the {\tt bilby dynesty} \cite{Ashton:2018jfp, Speagle:2019ivv} to implement Bayesian inference. We have also explored the impact of the GW signal intensity on the robustness of network inference. The results indicate that the SNR of the GW signal has varying degrees of impact on the network's ability to perform source parameter inference. Furthermore, we consider the influence of glitches' intensity on the robustness of source parameter inference when utilizing the normalizing flow network. Meanwhile, we conduct a detailed study on the impact of three key features of non-stationary noise \cite{Canton:2013joa}, on the robustness inference using the normalizing flow. We find that it is possible to determine the feasibility of obtaining reasonable results by assessing the robustness of the inference considering the characteristics of non-Gaussian noise.
%We find it is possible to determine whether reasonable results can be obtained based on the robustness of the inference by considering the characteristics of non-Gaussian noise. 
%We also explored whether the of GW source parameters can impact the robustness inference of normalizing flow network. 
%frequency $f$, quality factor $Q$, and amplitude $h_0$

The results demonstrate that the normalizing flow network can provide reliable posterior parameter inference for GWs contaminated by unknown glitches without undergoing deglitching.
%Without undergoing deglitching, the normalizing flow network can directly provide a reliable posterior inference of parameters for GW contaminated by unknown glitches. 
When GW data are contaminated by unknown noise, our model is capable of delivering swift and reasonable parameter inference, albeit with a difference in magnitude. This provides a reasonable prior for subsequent more accurate processing. When extended to binary neutron star (BNS) mergers, this method is also expected to enable the rapid localization of EM counterparts.

This paper is structured as follows. In Section~\ref{sec2}, the methodology used in the work is described in detail. Some of the work's conclusions and discussions are presented in Section~\ref{sec3}. Finally, we present the main conclusions of our work in Section~\ref{sec4}.

\section{Methodology}\label{sec2}
\subsection{Normalizing flow network architecture}\label{sec2.1}

%Also neural posterior estimation (NPE) \cite{rezende2015variational,NIPS2016_6aca9700,lueckmann2017flexible,greenberg2019automatic} will be used in the subsequent analysis process. The core idea of NPE is to generate a large number of simulated datasets and related parameters \cite{Dax:2021tsq}. These datasets are then used to train neural networks in normalizing flows, enabling the computation of posterior distribution.

The normalizing flow is a deep learning-based generative model for learning complex probability distributions \cite{rezende2015variational}. Its core idea is to transform a simple probability distribution into a complex probability distribution through a series of reversible transformations. Neural posterior estimation \cite{papamakarios2016fast} involves training a neural network to approximate the transformation process of this complex posterior distribution, thereby achieving an approximate estimation of the posterior distribution. In the inference process, the normalizing flow generates samples by applying a series of reversible transformations to the base distribution. The normalizing flow network learns the parameters of these transformations to map the base distribution to a posterior distribution corresponding to the real data distribution.

%The normalizing flow also has a wide range of applications in generative modeling, where they can be used for tasks such as sample generation, probability density function calculation, and data reconstruction. As the normalizing flow is reversible, both the generation and inference processes are solvable, making it a powerful and flexible modeling technique.
%Between coupling transformations, the order of the sample components is randomized to ensure that all components are fully transformed by the transformation sequence, specifically, we use the rational quadratic spline coupling transformation.

The normalizing flow network consists of a series of coupling transformations, each of which is transformed according to the input data and the output of the previous layer of the neural network, allowing the predicted distribution to gradually approximate the posterior distribution. We also employed neural spline flow \cite{durkan2019neural}, which utilizes monotonic rational quadratic spline coupling transformations. Between coupled transformations, sample components were inverted to ensure that the transformation sequence specifically and completely transformed all components.

%Between coupling transformations, the order of the sample components is randomized to ensure that the transformation sequence specifically fully transforms all components.

The normalizing flow \cite{rezende2015variational, kingma2016improved,chen2016variational, papamakarios2017masked} can be expressed by the following formula
\begin{equation}
q(\theta \mspace{-5mu} \mid \mspace{-5mu} s)=\pi\left(f_s^{-1}(\theta)\right)\left|\operatorname{det} J_{f_s}^{-1}\right|,
\end{equation}
where $f_s$ represents the invertible transformation of the normalizing flow \cite{rezende2014stochastic}, and $s$ represents the input data strain to the network. We select the base distribution $\pi(u)$ in a way that facilitates easy sampling and evaluation of its density \cite{Green:2020dnx}. In our case, we consistently choose the standard multivariate normal distribution with the dimension $D$ matching that of the sample space. We approximate the posterior distribution of GW $p(\theta|s)$, by utilizing the normalizing flow $q(\theta|s)$, aiming to obtain the posterior distribution from the parameter $\theta$.

\begin{figure*}[!htp]
\includegraphics[width=0.9\textwidth]{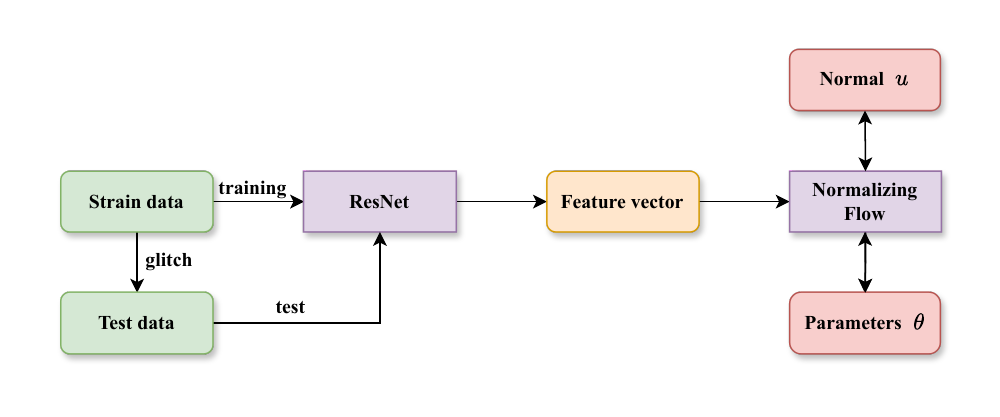}
\caption{\label{fig1} The workflow of the normalizing flow network. The input strain data contain simulated GW signals and Gaussian noise. The strain data is passed through an identical residual net (ResNet) model, and then the feature vectors are extracted from the ResNet model. Then it proceeds through the normalizing flow to obtain the parameter posterior distributions. The above steps are repeated by injecting glitches into the strain data as a test set.}
\end{figure*}

Our architecture for the normalizing flow is quite similar to the one described in Ref.~\cite{Sun:2023vlq}, except for data fusion. Our network architecture combines a one-dimensional ResNet-50 network \cite{he2016deep} for feature extraction with the normalizing flow network for estimating the parameters of the posterior distribution. The overall structure of the working principle is illustrated in Fig.~\ref{fig1}. The data strain input to the network first undergoes the ResNet neural network, then the feature vectors extracted from the ResNet model. Finally, the feature vectors are processed by the normalizing flow to obtain the GW waveform as well as the parameter inference.

Unlike the parameters used in Ref.~\cite{Sun:2023vlq}, the network parameters in this work are shown in Table~\ref{tab1}. For networks corresponding to different numbers of detectors, their network parameters remain the same, except that the number of channels in the network corresponds to the number of detectors. We also experimented with model complexity reduction and regularization techniques to enhance the robustness of the network but found that they had minimal impact on the results.

During the training process, we utilize the AdamW optimizer \cite{loshchilov2018decoupled} with a learning rate set to 0.0001, a batch size of 200, and a learning rate decay factor of 0.99. To ensure the validity of the normalizing flow network predictions and prevent overfitting, we regenerate 10,000 new data points at the beginning of each epoch and employ 50 cycles of early stopping. The training of this network was conducted on an NVIDIA RTX A6000 GPU with 48 GB of memory.

\begin{table}
\caption{The key hyperparameters of the normalizing flow.}\label{tab1}
\centering
\setlength\tabcolsep{18pt}
\renewcommand{\arraystretch}{1.5}
\begin{tabular}{cc}
\hline \hline Hyperparameter & Value \\
\hline Flow Steps & $9$ \\
Hidden Layer Size & $4096$ \\
Transform Blocks & $7$ \\
Activation & ReLU \\
Batch Norm & False \\
Bins & $32$ \\
\hline \hline
\end{tabular}
\end{table}

\subsection{Data set}\label{sec2.3}
In terms of dataset preparation, the data used during the training phase consists of simulated Gaussian noise and simulated GW signals. The signals used are from the GW signals simulated by the SEOBNRv4\_opt model \cite{Bohe:2016gbl}, and they are selected in the range of SNR between 8 and 32.
The noise used is a simulated Gaussian noise generated using {\tt Pycbc} \cite{Biwer:2018osg,DalCanton:2014hxh} with a frequency range of $20 - 4096$ Hz, which represents the high-pass frequency and sampling frequency respectively. {This also implies a Nyquist frequency of 2048 Hz.} To obtain the noise sensitivity curves ${S_{n,k}(f) }$ for different detectors, we utilize the aLIGOZeroDetHighPower, AdvVirgo, and KAGRA methods from the pycbc.psd library \cite{alex_nitz_2022_6912865,Sellers:2022fyt}. 
%After generating the Gaussian noise, we inject signals with the specific parameter distribution shown in Table~\ref{tab2} and whiten data. 
After generating Gaussian noise, we inject a signal between 0.7 s and 0.9 s. The specific parameters distribution of the signal are shown in Table \ref{tab2}, and the simulated GW observation data are whitened. Finally, a 2-second dataset is generated.

Here, for a signal $h$, the SNR is defined as
\begin{equation}
\rho = (\boldsymbol{\tilde{h}} \mid \boldsymbol{\tilde{h}} )^{{1}/{2} },
\end{equation}where $\boldsymbol{\tilde{h}}$ is the frequency domain GW waveform considering the detector network including $N$ independent detectors and can be written as $\boldsymbol{\tilde{h}}=\left [  \tilde{h}_1, \tilde{h}_2,\cdot \cdot \cdot,\tilde{h}_k,\cdot \cdot \cdot,\tilde{h}_N\right ] $. Here $\tilde{h}_{k}$ is the frequency domain GW waveform of the $k$-th detector. The inner product is defined as \cite{Jin:2023sfc}
\begin{equation}
(\boldsymbol{\tilde{h}}\mid\boldsymbol{\tilde{h}}) = \sum_{k=1}^{N} (\tilde{h}_{k}\mid \tilde{h}_{k} )=\sum_{k=1}^{N} 4\int_{f_{\rm min}}^{f_{\rm max}} \frac{\tilde{h}_{k}(f) \tilde{h}_{k}^{\ast }(f)  } {S_{n,k}(f) } \mathrm{d}f \label{eq:detecsnr},
\end{equation}
and ${S_{n,k}(f) }$ is the sensitivity curve function of the $k$-th detector \cite{Owen:1998dk}. In the above expression, $f_{\rm min}$ is the high-pass frequency, and $f_{\rm max}$ is the Nyquist frequency. Also, we calculate the glitch SNR in the same way.
%To ensure the rationality of the normalizing flow network's prediction results and prevent overfitting, it is common practice to regenerate 10,000 new data points at the start of each epoch.

\begin{figure}[!htp]
\centering
\includegraphics[width=0.5\textwidth]{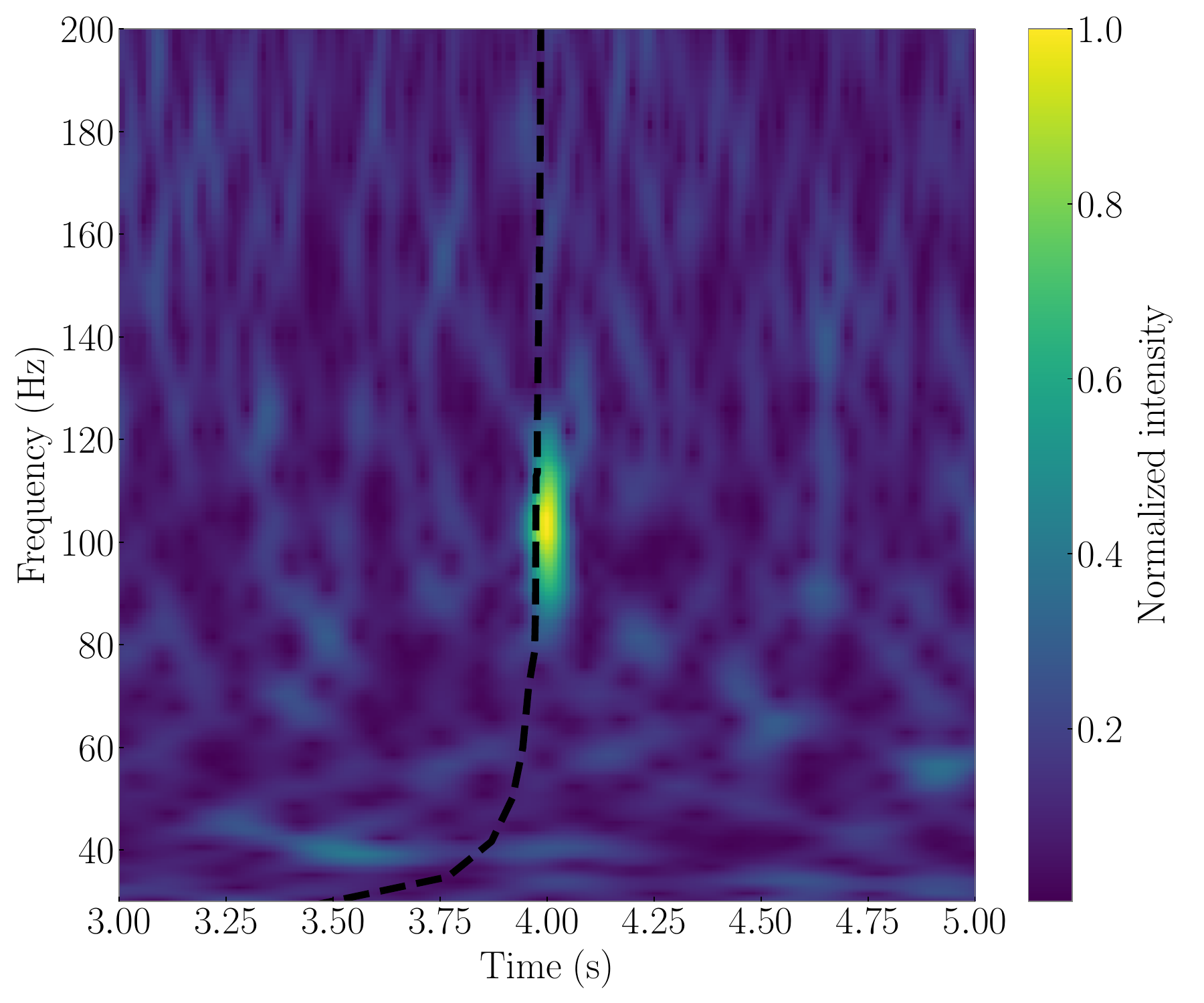}
\centering \caption{\label{fig2}Time-frequency spectrogram for a simulated GW signal coincident with a glitch. The black dashed track shows the inferred time-frequency evolution of the GW signal. The light spot is the approximate location of the simulated glitch.}
\end{figure}

\begin{table}
\caption{Priors of simulated GW waveform parameters. Note that other parameters not mentioned are set to zero for simplicity.}\label{tab2}
\centering
\setlength\tabcolsep{18pt}
\renewcommand{\arraystretch}{1.5}
\begin{tabular}{cc}
\hline \hline Parameter & Uniform distribution \\
\hline Chirp mass & $\mathcal{M}_{\rm c}\in[25.0,62.5]~ M_{\odot}$ \\
Mass ratio & $q\in [0.5,1]$\\
Right ascension & $\alpha\in[0, 2\pi]~{\rm rad}$ \\
Declination & $\delta\in[-\pi/2,\pi/2]~{\rm rad}$ \\
Polarization angle & $\psi\in[0,2 \pi]~{\rm rad}$ \\
Luminosity distance & $d_{\rm L}\in[500,5000]~\mathrm{Mpc}$\\
\hline Signal to noise ratio & ${\rm SNR}\in[8,32]$ \\
\hline \hline
\end{tabular}
\end{table}

In the test dataset, we utilize simulated Gaussian noise and simulated GW signals, injecting simulated glitches into them, as illustrated in Fig.~\ref{fig2}. The glitches we use are modeled using sine-Gaussian functions \cite{Blackburn:2008ah}, which cover most of the types of glitches currently observed by the LIGO, Virgo, and KAGRA networks. {Note that when injecting glitches into detectors, we selected the LIGO Livingston detector due to its higher sensitivity and higher occurrence frequency of glitches, as random glitches typically occur in only one detector.} We examine a sine-Gaussian glitch \cite{Powell:2015ona}, which is defined as
\begin{equation}
h(t)=h_{0}\sin (2\pi f_{0}(t-t_{0} ))\exp ({- \frac{(t-t_{0} )^{2} }{2\tau ^{2}} } ),
\end{equation}where $\tau =Q/\sqrt{2} \pi f_{0}$ is the time duration. Here $h_{0}={\rm hrss}/\sqrt{\tau}$, where $\rm hrss$ is the root sum squared amplitude of the glitch, $Q$ is the dimensionless quality factor, $t_{0}$ and $f_{0}$ are the location of the sine-Gaussian in time and frequency respectively.
%We can define the $t_{0}= 0$, so $t$ represents the time delay between the sine-Gaussian glitch and the inspiral trigger. 
We set $t_{0}=0.8$ s to simulate the scenario where the signal and glitch overlap {(the underlying rationale is provided in Appendix \ref{appendix:A})}. 
The selected parameters range of glitches are shown in Table~\ref{tab3}. 

\begin{table}
\caption{Parameters used to prepare the simulated glitches set.}\label{tab3}
\centering
\setlength\tabcolsep{18pt}
\renewcommand{\arraystretch}{1.5}
\begin{tabular}{cc}
\hline \hline Parameter & Value \\
\hline $f_{0}$ & $50-300~ {\rm Hz}$ \\
$Q$ & $2-20$ \\
$h_0$ & $4\times 10^{-22}-1\times 10^{-21}$ \\
\hline \hline
\end{tabular}
\end{table}

\subsection{Analytical approach}\label{sec2.4}

%The Kolmogorov-Smirnov (KS) \cite{Veitch:2014wba} sample test is a non-parametric statistical test used to compare two samples or the difference between a sample and a theoretical distribution. 

The Kolmogorov-Smirnov (KS) test \cite{Lopes2011} is a statistical method that can be used to test the goodness of fit between a sample distribution and a reference distribution. In the context of GW parameter estimation, the KS test can be employed to assess the agreement between the estimated parameter distribution and a theoretical or expected distribution \cite{Sidery:2013zua,Veitch:2014wba,Biwer:2018osg,Thrane_Talbot_2020}. The KS test first calculates the cumulative distribution functions (CDFs) of the two distributions. Then it evaluates the distance $D_{\rm {n}}$ between the two distributions based on the CDFs, which can be written as
\begin{equation}
D_{n} = \max \mid F_{n}(x)-F(x)\mid,
\end{equation} 
where $F_{n}(x)$ is a specified continuous distribution function from a random sample that a set of random variables $X_i$, $i=1,...,n$. $F_{n}(x)$ and $F(x)$ respectively represent the CDFs of the two distributions under validation. The resulting $D_{n}$ is also known as the KS statistic. The p-value, calculated based on the KS statistic, is commonly employed as a straightforward measure to assess the similarity or dissimilarity between two distributions. The p-value is commonly compared to the significance level. If the p-value is less than the significance level, it indicates that the two distributions are considered statistically different. Here, the significance level has taken a value of 0.05. If the p-value obtained from the KS test is less than 0.05, it indicates that the two distributions are statistically different, and the hypothesis of their similarity can be rejected.

In this work, we compare two distributions: the ideal distribution and the one-dimensional distribution generated by the normalizing flow network. Since the percentile should be uniformly distributed between 0 and 1, the ideal distribution is uniform. The other distribution consists of the probabilities corresponding to the confidence intervals in which the true value of each sample lies among the simulated 200 samples. Our initial hypothesis assumes that the two distributions are the same distribution, which means the posterior inference of the parameters given by the normalizing flow network is reasonable. Then the p-value is compared with the significance level of 0.05; if the p-value is less than 0.05, we consider the original hypothesis invalid.

The Jensen–Shannon divergence (JSD) is used to compare the difference between two probability distributions and measure the similarity of two distributions \cite{61115,rao1987differential}. It is achieved by taking a weighted average of the probability density functions of the two distributions, given by
\begin{equation}
{\rm JSD}(p\parallel q )=\frac{1}{2} {\rm KLD}(p \parallel \frac{p+q}{2} )+\frac{1}{2} {\rm KLD}(q\parallel \frac{p+q}{2} ),
\end{equation}
where Kullback-Leibler divergence (KLD) \cite{10.1214/aop/1176996454,10.1214/aoms/1177729694} is defined as
\begin{equation}
{\rm KLD}(p\parallel q)=\int_{-\infty }^{+\infty } p_{(x)}\ln_{}{} \frac{p_{(x)} }{q_{(x)} } \mathrm{d}x.
\end{equation}
The KLD can also be used to measure the difference between two probability distributions, but the KLD is asymmetric, so calculating the KLD between two distributions $p$ and $q$ may yield different results than calculating the KLD between $q$ and $p$. To overcome the asymmetry of KLD, JSD is more symmetrical and smooth. JSD values range from 0 to 1, making explaining and comparing differences between different distributions easier. The JSD value ranges from 0 to 1, where a JSD value of 0 indicates identical distributions and a JSD value of 1 represents the maximum difference between the distributions \cite{Romero-Shaw:2020owr}. This work uses the JSD criterion to measure the difference between the parameter distribution inferred by the normalizing flow network after being contaminated by glitches and the ideal parameter distribution.

\section{Results and discussion}\label{sec3}

%We present our results in three subsections where we discuss what effect different detector numbers, GW parameters, and non-stationary noise have on robustness inference.

\begin{figure*}[!htp]
\centering
\includegraphics[width=0.8\textwidth]{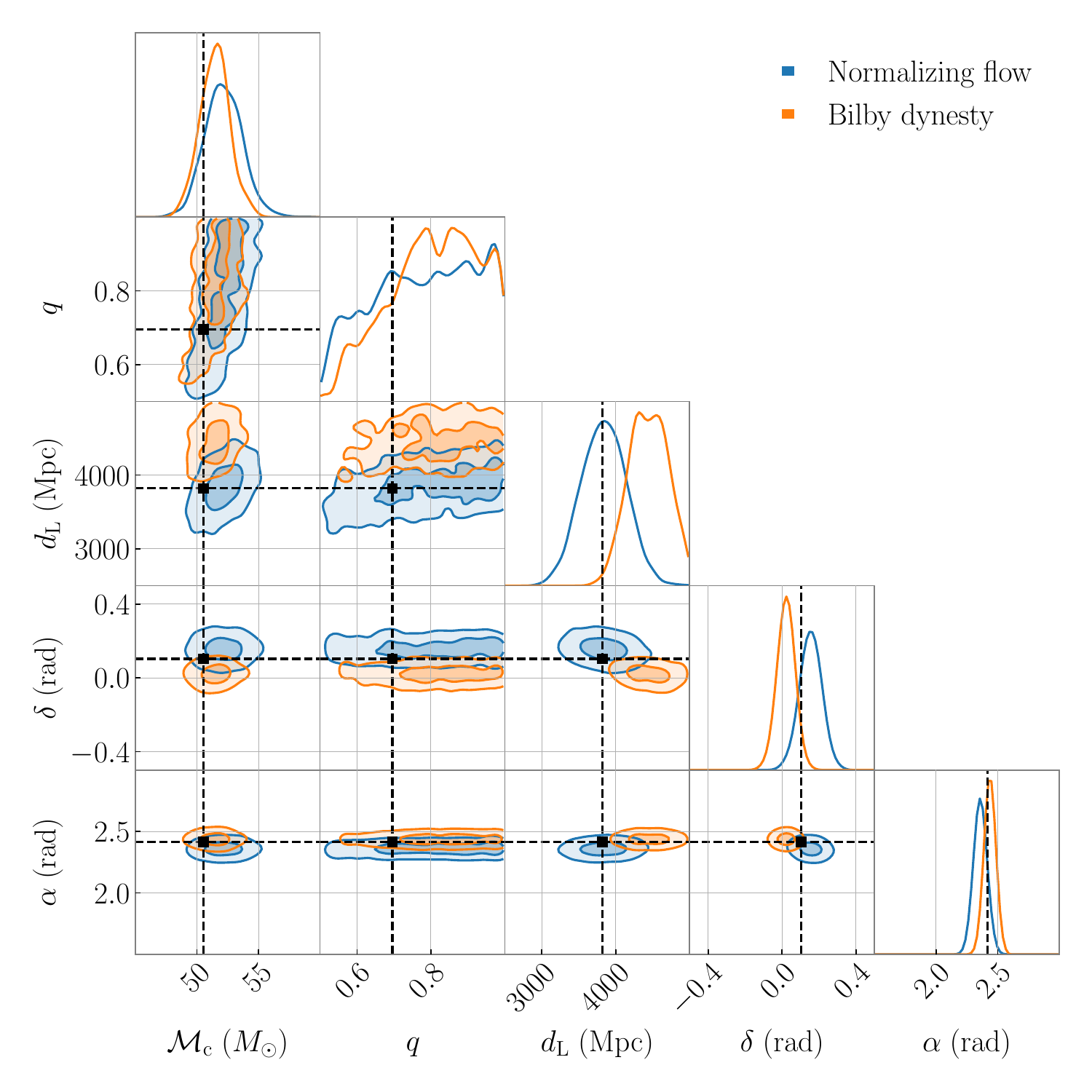}
\centering \caption{\label{fig3}Parameter posterior corner plot of injected signal in Gaussian noise with glitch SNR of 8. Here the blue and orange lines show the inference of the joint posterior distribution of the parameters by the normalizing flow and {\tt bilby dynesty}, respectively. One-dimensional histograms of the posterior distribution of each parameter in both methods are plotted along the diagonal. The injection values of the simulated signal are indicated by the black vertical and horizontal lines.}
%Light grey and dark grey areas indicate 2$\sigma$ and 1$\sigma$ confidence intervals respectively.
\end{figure*}

Before proceeding with further analysis, we conducted tests using the normalizing flow to assess the rationality of parameter inference in the presence of glitches. We injected glitches into the simulated GW data and performed parameter inference using the normalizing flow and {\tt bilby dynesty} respectively. The results are shown in Fig.~\ref{fig3}. It can be seen that the parameter posterior distribution obtained by the normalizing flow is closer to the injection values of parameters than the results inferred by the {\tt bilby dynesty}. This shows that the normalizing flow is superior to the conventional {\tt bilby dynesty} for processing GW data contaminated by glitches, which is the basis of our subsequent work.

\begin{figure}[!htp]
\centering
\includegraphics[width=0.5\textwidth]{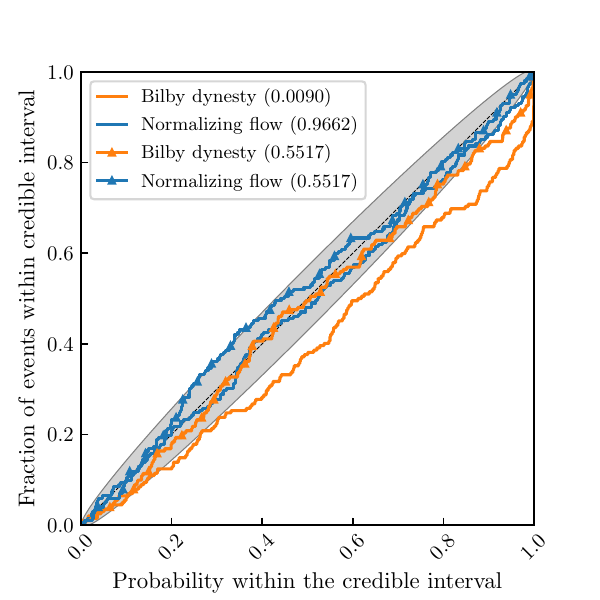}
\centering \caption{\label{fig4}Cumulative distribution of the quantile of which the true value lies for marginalized distribution. The curves were constructed using the 200 samples and the dashed black diagonal line indicates the ideal result. Shaded areas indicate 95\% confidence intervals from the ideal distribution with the same number of events as the injection samples. The legend shows p-values obtained by the normalizing flow and {\tt bilby dynesty}. {The triangular line represents p-values without glitches.}}
\end{figure}

We simulate 200 events to validate the robustness of the normalizing flow network's inference for arbitrarily selected parameters. By injecting 200 simulated GW waveforms, along with the added glitches, we measured the frequency of the true parameter over a range of confidence levels. {We also simulated 200 events without glitches.} We calculate the cumulative probability of each parameter posterior's percentile at the injection values of parameters, then plot the CDF for the joint distribution of the source parameters in Fig.~\ref{fig4}. Ideally, the injected CDF of the parameters should exhibit a diagonal pattern, as the CDF represents the cumulative sum of individual event distributions. The proximity of the curve to the diagonal indicates the accuracy of parameter inference. A curve that closely aligns with the diagonal signifies superior performance in accurately estimating the parameters. The legend displays p-values obtained from the posterior distributions for each method.
%We additionally conducted simulations on 200 events without the inclusion of glitches, thereby assessing the performance of both methods under Gaussian noise.

Based on our assumption that p-values less than 0.05 are deemed unreasonable for the model to support true posterior inference, the p-value provide a quantitative measure of the agreement between the normalizing flow model cumulative distribution and the expected theoretical distribution. {With only Gaussian noise, both the normalizing flow and {\tt bilby dynesty} methods perform well.} But the normalizing flow employed in our study demonstrates superior performance compared to the {\tt bilby dynesty} when it comes to accurately learning the true posterior distribution, particularly in scenarios involving glitch-contaminated GW data.

%In the scenario with only Gaussian noise, both the normalizing flow and the {\tt bilby dynesty} method perform admirably.

\begin{figure}[!htp]
\centering
\includegraphics[width=0.5\textwidth]{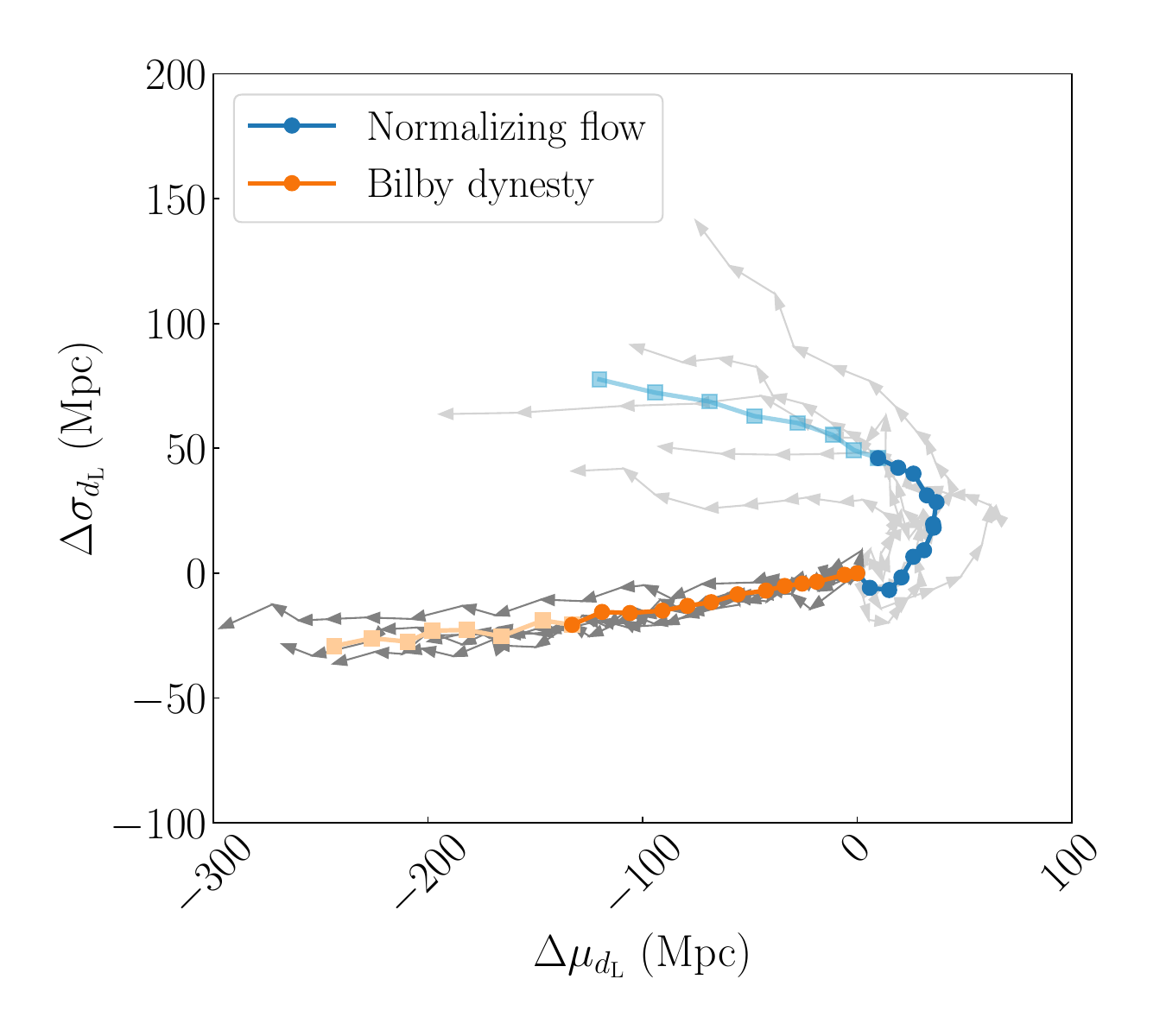}
\centering \caption{\label{fig5}Variation of the difference in the mean and standard deviation of inferred GW source parameter $d_{\rm L}$ with increasing glitch SNR. The dark and light gray lines represent the results from five independent experiments for the normalizing flow and the {\tt bilby dynesty}, respectively. The arrows represent the direction of increasing glitch SNR. The blue and orange lines represent the average results from five independent experiments. {Circles and squares indicate glitch SNR below and above 8, respectively.}}
\end{figure}
%The dark and light shades of the blue and orange lines respectively represent glitch SNR less than 8 and greater than 8. 

From Figs.~\ref{fig3} and \ref{fig4}, it can be observed that our normalizing flow model outperforms the {\tt bilby dynesty} in posterior parameter inference. We investigated the reasons behind the superior performance of the neural network, which are demonstrated in Fig.~\ref{fig5}. To generate different sets of data, we generate different Gaussian noises by changing the random seed and then injecting glitches and signals into them. We selected five different sets of Gaussian noise data, with each set having glitch SNR ranging from 0 to 19. Each point on the same line consists of 200 simulated GW data, where these events have the same GW signal and Gaussian noise, but different injected glitches. We plotted the differences between the mean and standard deviation of the fitted posterior distributions of parameters obtained by the normalizing flow network and the Bayesian inference method for different glitch SNRs. These differences were calculated by subtracting the results obtained at a glitch SNR of 0.

We selected the $d_{\rm L}$ parameter for demonstration purposes. As the glitch SNR increases, the mean and standard deviation obtained by the normalizing flow tend to be centered around the injected values when the glitch SNR is relatively low. However, they gradually deviate from the injected values as the glitch SNR increases. In contrast, the {\tt bilby dynesty} continues to deviate from the injected values persistently as the glitch SNR increases. This also demonstrates the results above, specifically in Fig.~\ref{fig3}, where the normalizing flow provides a wider range of posterior distributions as the injected values increase, while the {\tt bilby dynesty} shows more deviation from the injected values. Indeed, this also highlights the robustness of the normalizing flow in handling glitches. The normalizing flow exhibits a certain level of stability and resilience in the presence of glitches, as evidenced by its ability to provide posterior distributions that are closer to the true values compared to the {\tt bilby dynesty}.

\subsection{Different numbers of detectors}\label{sec3.1}

Indeed, with the increase in the number of future GW detectors and improvements in their sensitivity, it is expected that the number of events contaminated by glitches will also significantly increase. This will pose even greater challenges for GW data processing. In principle, increasing the number of detectors will result in a lower proportion of contaminated data considering only one detector data contaminated by glitches. Indeed, in the case of the same SNR of the detector networks, according to Eq.~(\ref{eq:detecsnr}), the SNR for each detector will decrease. Therefore, the intensity of the detected GW signals by a single detector will also decrease, resulting in a higher relative intensity of glitches compared to the signal, as compared to a network with fewer detectors. In this section, we investigate the impact of the number of detectors on our normalizing flow network capability to infer parameters. 

\begin{figure}[!htp]
\centering
\includegraphics[width=0.5\textwidth]{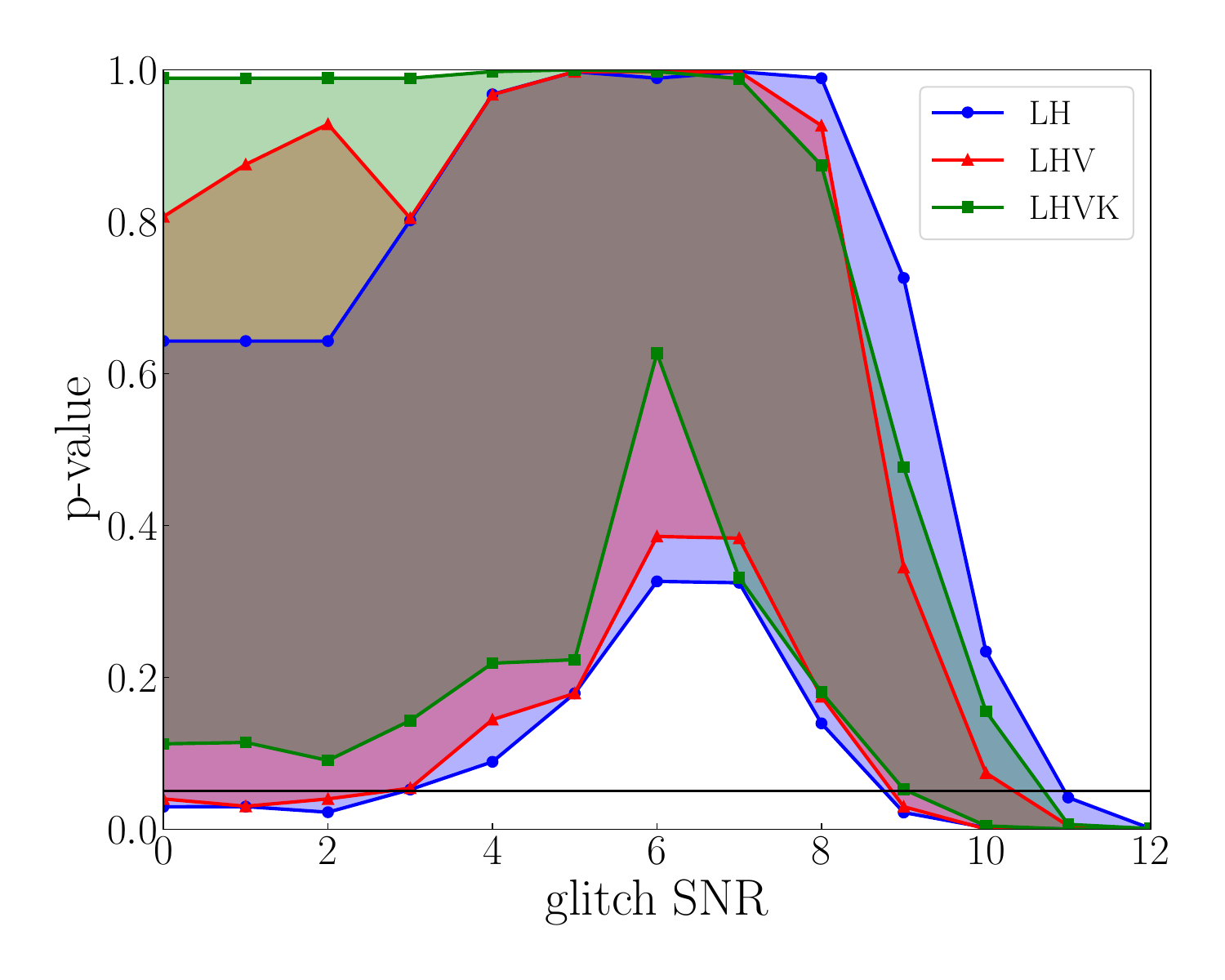}
\centering \caption{\label{fig6}P-values of different glitch intensities with different number of detectors. The legend provides information about the colors and labels associated with the curves representing different detectors. Specifically, ``LH'' represents the LIGO network, ``LHV'' represents the LIGO and Virgo network, and ``LHVK'' represents the LIGO, Virgo, and KAGRA network \cite{KAGRA:2013rdx}. The black line represents a p-value of 0.05.}
\end{figure}

%and the reasons are shown in Fig.~\ref{fig5}
Figure~\ref{fig6} illustrates p-values obtained parameter inference when glitches were injected into the GW data from different numbers of detectors, respectively. It depicts the distribution of p-values error bars for 20 sets of 200 GW data with varying detectors as the injected glitch SNR increases. These error bars are derived from the $1 \sigma$ confidence intervals of the values provided by each group. The curves of the detectors exhibit an upward trend and then a downward trend. The normalizing flow can exhibit overconfident posterior approximations, leading to underestimated parameter variances \cite{hermans2022trust}. However, as shown in Fig.~\ref{fig5}, it can be observed that when the glitch SNR is low, the mean value of the parameters does not change significantly, while the standard deviation increases noticeably. This behavior partially offsets the excessive confidence in the normalizing flow, leading to an increase in p-values in parameter inference calculations when glitches are absent. Therefore, when the glitch is small, the p-value will tend to increase. {Since the p-values are bounded at 1, the expansion of the upper bounds of the error bars is constrained, while a more pronounced rise is observed in the lower bounds. This interplay between the inherent overconfidence of the network and the additional variance introduced by glitches leads to the tightening effect in Fig.~\ref{fig6} when the glitch SNR reaches 6.}
%As a result, the plot exhibits a trend of initially increasing.
%However, as shown in Fig.~\ref{fig5}, it can be observed that when the glitch SNR is low, the standard deviation of the parameters tends to be larger.
%However, as shown in Fig.~\ref{fig5}, it can be observed that when the glitch SNR is low, the standard deviation of the parameters tends to be larger.

Notably, the lower and upper limits of the error curves for the four detectors demonstrate a smoother downward trend compared to the error curves of the two or three detectors. The intersection point of the lower limit of error for these curves with a p-value of 0.05 occurs at a glitch SNR of approximately 9. This observation highlights that beyond a glitch SNR of 9, the inference results of the network's parameter estimation become unreliable for any number of detectors. The presence of glitches severely affects the accuracy of the parameter inference in such cases. This result indicates that as the intensity of the glitch increases, the influence of the glitch on different detector numbers gradually becomes more significant. Although there may be slight differences, overall, the impact is relatively consistent across the detectors.

\subsection{The SNR of gravitational wave signal}\label{sec3.2}

\begin{figure*}[!htp]
\centering
\includegraphics[width=1\textwidth]{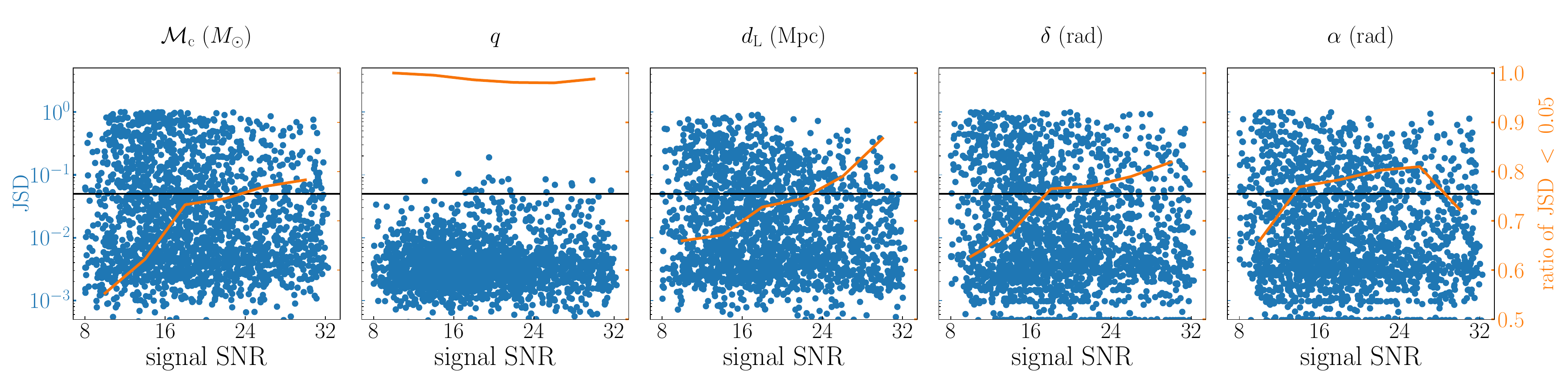}
\centering \caption{\label{fig7}JSD of GW parameter marginal distributions as a function of signal SNR between glitch SNR of 8 and without glitch. The blue scatter plot represents the distribution of JSD values for the marginal distributions of the parameters of the 2000 simulated GW events. The black line represents a JSD value of 0.05. The orange line represents the ratio of the number of events with JSD values less than 0.05 to the total number of events, for different signal SNR.}
\end{figure*}

In this section, we investigate the impact of signal SNR on the normalizing flow network's capability to facilitate GW parameter inference. By examining these metrics, we aim to gain insights into the normalizing flow network's performance in accurately estimating GW parameters.

We explore the effect of the SNR of signal on the robustness of the network, using a glitch with 8. For this purpose, we selected several parameters, namely, $\mathcal{M}_{\rm c}$, $q$, $d_{\rm L}$, $\delta$, $\alpha$. These parameters have a significant impact on GW signal simulation, parameter estimation, and sky localization. By considering these parameters in our analysis, we aim to comprehensively analyze their effects and gain a deeper understanding of their implications in GW studies. For quantitative comparison, we computed the JSD of the marginal distributions of the GW parameters between glitch SNR of 8 and without glitch, shown in Fig.~\ref{fig7}.

The blue portion depicts the distribution characteristics of JSD values for the marginal distributions of parameters among the 2000 simulated GW events. To highlight the trend more explicitly, an orange line is plotted to showcase the overall variation in the number of events with JSD values less than 0.05. From the orange line, it is evident that the five parameters of GW exhibit varying trends in terms of their JSD values. While the parameter $q$ demonstrates a relatively flat trend, the remaining four parameters $\mathcal{M}_{\rm c}$, $d_{\rm L}$, $\delta$, $\alpha$ show an increasing trend in the percentage of values with JSD values below 0.05, relative to the total number of values.

The black line indicates a JSD value of 0.05. Based on the assumption mentioned in the section, a JSD value below 0.05 indicates a difference between the two distributions that cannot be ignored. It also implies that the parameter inferences provided by the normalizing flow network are no longer reasonable. As signal SNR increases, the $\mathcal{M}_{\rm c}$, $d_{\rm L}$, $\delta$, $\alpha$ parameter inference results are affected by glitches, and the greater SNR, the less likely these four parameter inference results are to be affected by glitches. In other words, a stronger SNR enhances the robustness of the parameter inference process and reduces the impact of glitches on the accuracy of the inferred values for $\mathcal{M}_{\rm c}$, $d_{\rm L}$, $\delta$, $\alpha$. The reasons for the trend of $q$ parameter inference results are given in \cite{61115}. Due to the poor limiting accuracy of $q$, it shows that JSD itself is smaller in the same case with a large variance.

%\subsection{Non-stationary noise}\label{sec3.3}
{\subsection{The SNR of the glitch}\label{sec3.3}}

\begin{figure*}[!htp]
\centering
\includegraphics[width=1\textwidth]{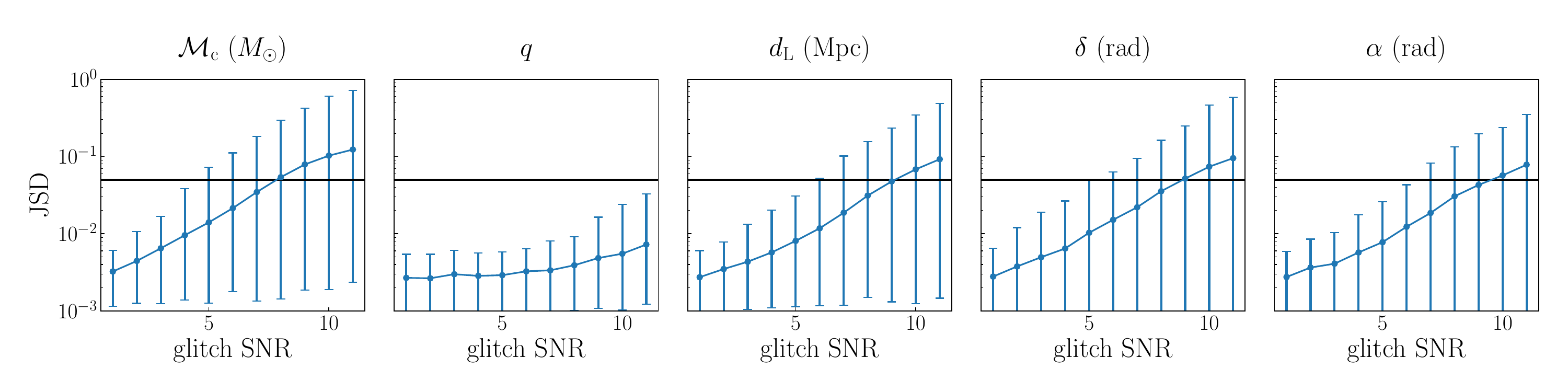}
\centering \caption{\label{fig8}JSD for GW parameter marginal distributions as different SNR of glitch between glitch-contaminated signals and Gaussian noise. The error bars on JSD represent the range of $90\%$ confidence interval. The black line represents a JSD value of 0.05.}
\end{figure*}

In this section, we explore the effect of the glitch's parameters and intensity on the robustness of the normalizing flow network. Despite the identification of sources for some glitches, a significant proportion of glitches remain of unknown origin. At the same time, due to the significant uncertainties associated with fault models, there are limitations to eliminating glitches through modeling techniques \cite{Ashton:2022ztk}. Here, we investigate the impact of glitches on parameter estimation. We aim to figure out the robustness of the normalizing flow network against unknown glitches, sequentially inferring parameters more quickly and accurately. 
%We explore the potential of utilizing the robustness of normalizing flow networks to mitigate the effects of these unknown glitches on parameter inference.

Here we investigate how the SNR of the glitches affects the ability of the network to infer GW parameters. Between the glitch-contaminated signals and Gaussian noise, we computed the JSD values for the marginal distributions of GW parameters at various SNRs of glitches. The results are presented in Fig.~\ref{fig8}, providing a graphical representation of the JSD values and enabling a comparative analysis of the parameter fringes under different glitch SNRs. The black line on the graph represents the JSD threshold of 0.05. Except for the $q$ parameter, whose JSD values are all below 0.05, the intersection of the error bars of the $\mathcal{M}_{\rm c}$, $d_{\rm L}$, $\delta$, $\alpha$ parameters with a JSD of 0.05 lies between glitch SNR between 4 and 7. Among the parameters, the intersection point between the error bar of parameter $\mathcal{M}_{\rm c}$ and JSD equal to 5 occurs at a glitch SNR between 4 and 5. In comparison to the other parameters, the posterior distribution of parameter $\mathcal{M}_{\rm c}$ is more susceptible to contamination from glitches.

\begin{figure*}[!htp]
\centering
\includegraphics[width=1
\textwidth]{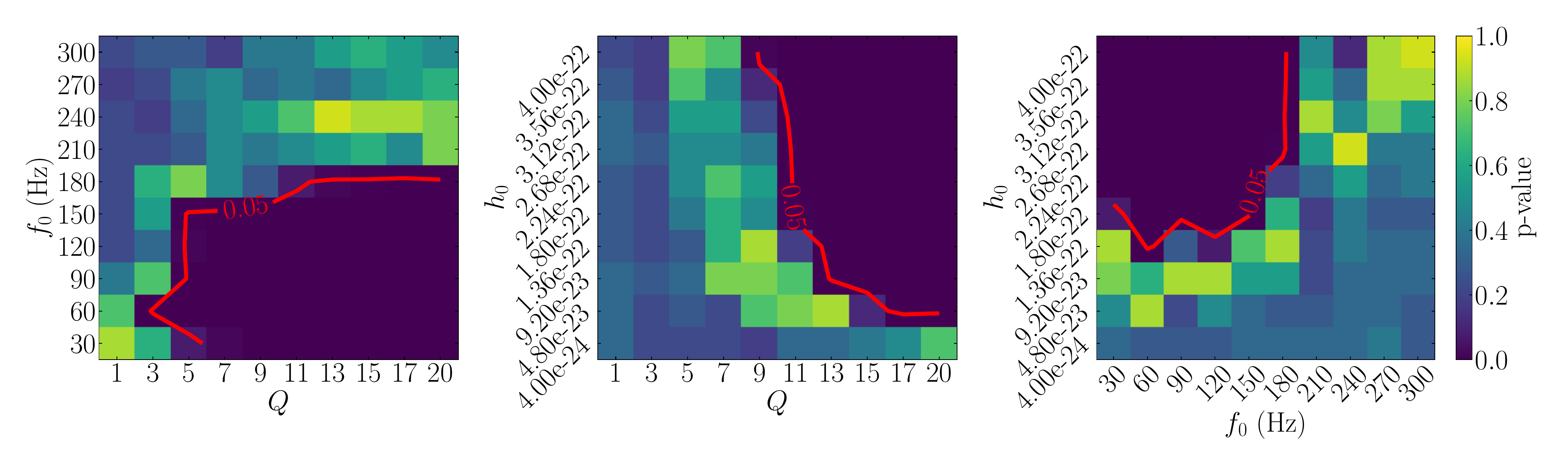}
\centering \caption{\label{fig9}The effect of glitch parameters on the inferred parameters of GW. Separate plots display the p-value distributions of the two-by-two combinatorial inference maps for the three fault parameters. The red line corresponds to a p-value of 0.05.}
\end{figure*}

The parameters of glitch that we explore are those used in the simulation of glitch: $f_0$, $h_0$ and $Q$. Figure~\ref{fig9} shows the p-values plot of the normalizing flow network affected by the glitch parameter, drawn according to the KS test. We vary the glitch's frequency, strength, and duration and ensure they are within a reasonable range. The horizontal and vertical coordinates on the graph are the two combinations of the glitch's parameters. According to our assumption that the p-value is less than 0.05, the inference of network parameters is unreasonable. It can be seen in Fig.~\ref{fig9} that each parameter of glitch has an impact on the accuracy of the network's inference of GW parameters. Specifically, as the duration of the glitch increases, along with a decrease in frequency and an increase in amplitude, the adverse effect on the rationality of parameter inference becomes more significant. Taking into account the upper limit of the GW signals frequency range, which is 200 to 400 Hz, and the lower limit of 20 Hz, it is evident that the impact of glitches on the inference of parameter inference becomes increasingly significant as more glitches overlap in both time and frequency with the GW signals.

\section{Conclusion}\label{sec4}

When analyzing and processing GW signals, glitches are a significant factor affecting the inference of GW parameters and source inversion. Several methods exist to model and remove glitches, but these approaches are more effective when the glitches are strong. For glitches with a SNR below 7, identifying them based on their morphology in images is challenging \cite{bahaadini2018machine}. Consequently, removing these low-SNR glitches is difficult. To address this, we explore the removal of glitches by leveraging the inherent robustness of the normalizing flow network, without the need for explicit glitch modeling.

When GW data are contaminated by unknown non-Gaussian noise, it is possible to directly perform parameter inference using a normalizing flow network trained only on Gaussian noise. For unknown and unmodeled glitches, if the glitch SNR is within an appropriate range, the robustness of the normalizing flow network ensures reasonable parameter inferences. Additionally, we investigate the reasons behind the differences in the posterior distribution of source parameters observed between the normalizing flow and Bayesian inference methods, providing detailed explanations.

We also find that the parameters of the glitches significantly impact the robustness of parameter inference using the normalizing flow network, compared to the GW parameters. This highlights the importance of accurately estimating the characteristic parameters of glitches when dealing with data contaminated by them. When glitches and GW signals overlap in time and frequency, it can significantly affect the parameter inference of the normalizing flow network, reducing the network's robustness and making accurate inference challenging.

This study is the first to investigate the robustness of the normalizing flow network in the presence of non-stationary noise, demonstrating its practical feasibility for addressing unknown interference. Moreover, machine learning methods can provide prior support to traditional methods, accelerating the inference of GW signal parameters and improving accuracy to some extent. With future model improvements, we aim to extend this method to BNS systems, accelerating the search process for electromagnetic counterparts.

\begin{acknowledgments}
This research has made use of data or software obtained from the Gravitational Wave Open Science Center (gwosc.org), a service of LIGO Laboratory, the LIGO Scientific Collaboration, the Virgo Collaboration, and KAGRA. We thank Shang-Jie Jin and Wan-Ting Hou for the helpful discussions. This work was supported by the National SKA Program of China (Grants Nos. 2022SKA0110200 and 2022SKA0110203), the National Natural Science Foundation of China (Grants Nos. 12473001, 11975072, 11875102, and 11835009), and the National 111 Project (Grant No. B16009).

\end{acknowledgments}

\begin{figure}[!htp]
\centering
\includegraphics[width=0.5\textwidth]{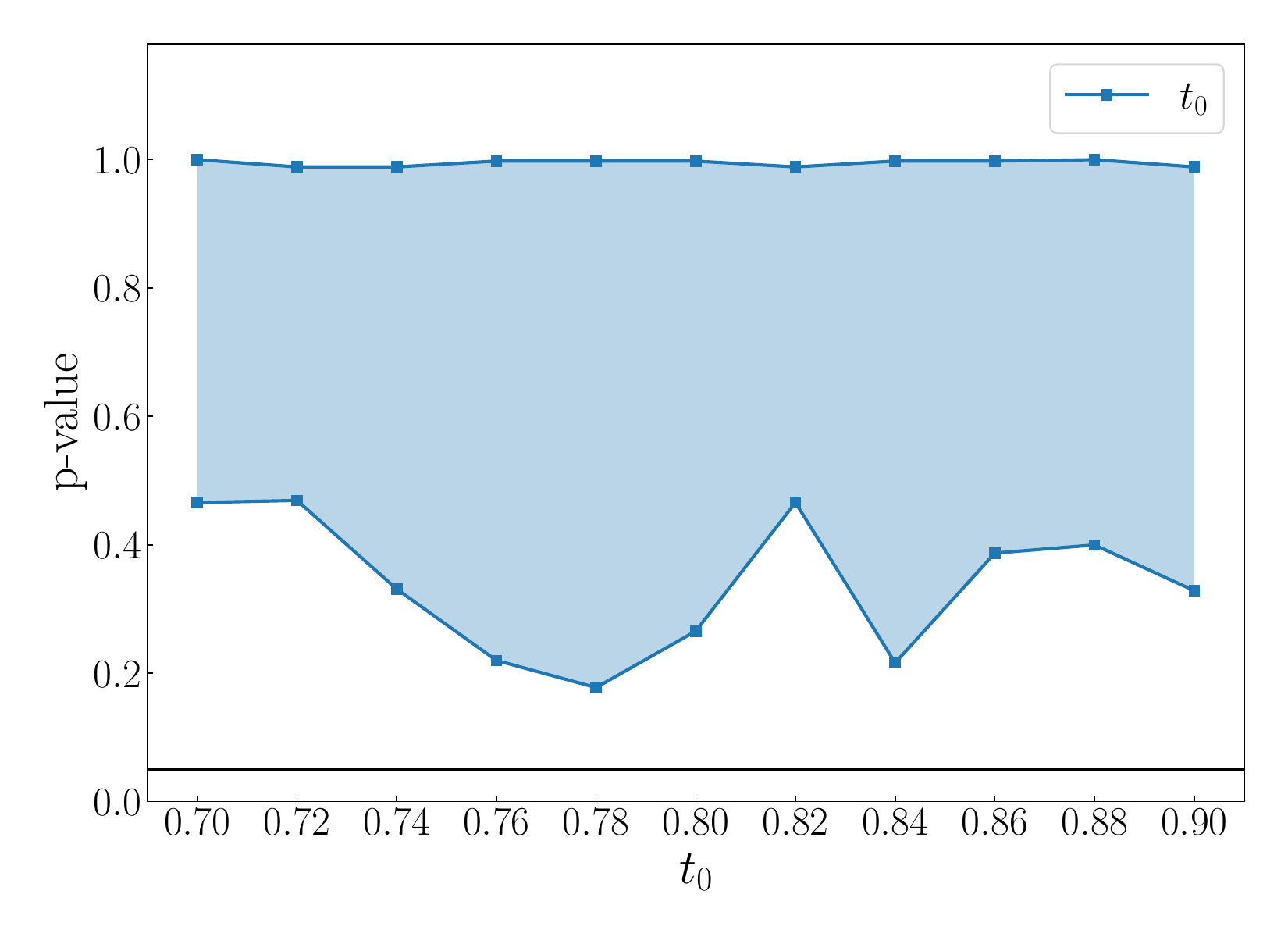}
\centering \caption{\label{fig10}{The effect of $t_{0}$ on the inference of network parameters. The variation of p-values over different $t_{0}$. The blue error bar represents the $1 \sigma$ confidence intervals for 20 sets of results. The black line represents a p-value of 0.05.}}
\end{figure}

\begin{figure}[!htp]
\centering
\includegraphics[width=0.5\textwidth]{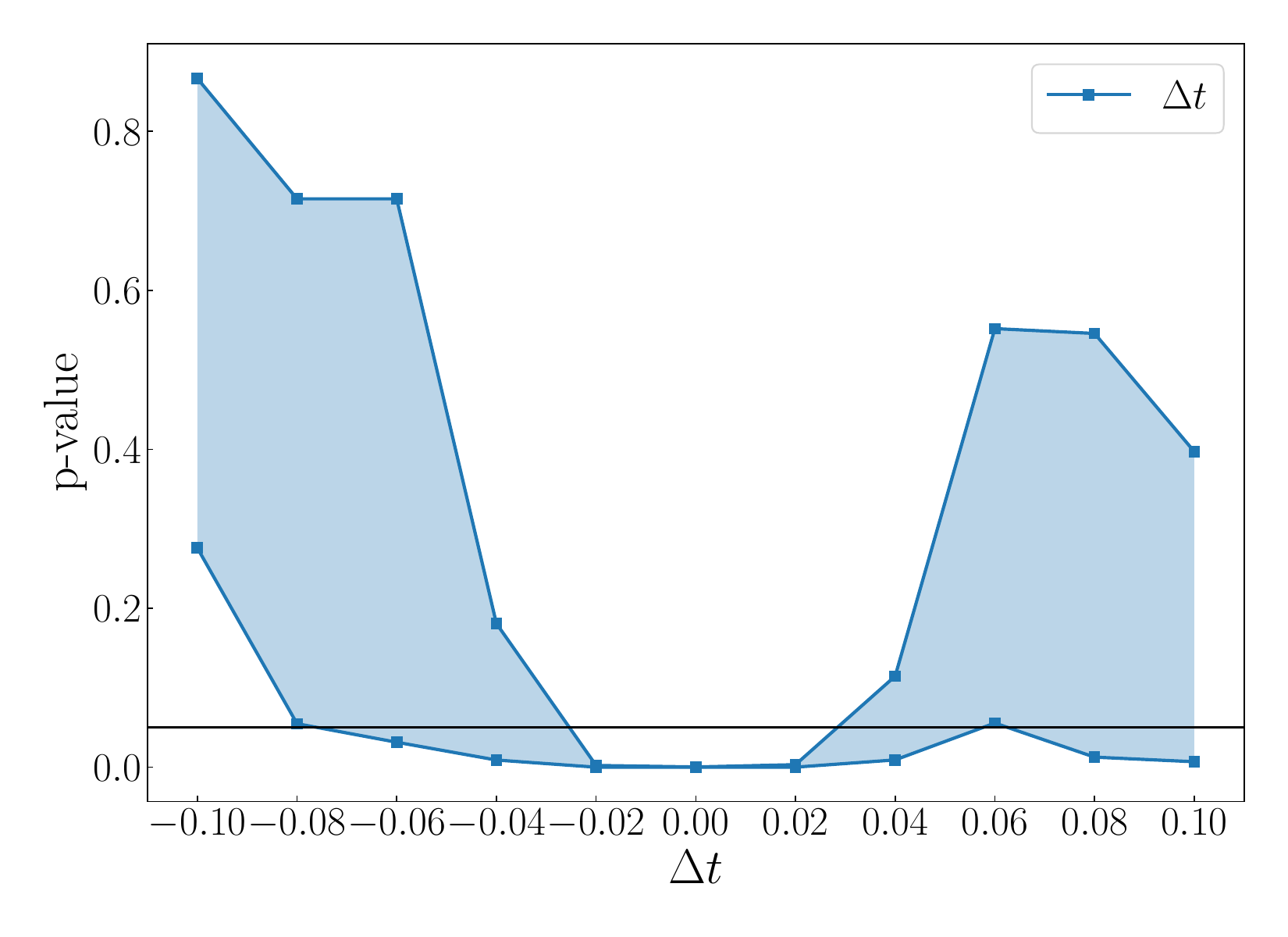}
\centering \caption{\label{fig11}{The effect of $\Delta t$ on the inference of network parameters. The p-values of parameter inference under different $\Delta t$. The blue curves illustrate the $1 \sigma$ confidence bounds. The black line represents a p-value of 0.05.}}
\end{figure}

\appendix 
\section{{Glitch time parameter}}
\label{appendix:A}

%Here, we analyze the impact of glitch injection time $t_{0}$ and the time interval between glitch and GW signal on the inference of network parameters (hereinafter referred to as $\Delta t$).
{In this Appendix, we examine the impact of glitch central time $t_{0}$ and its interval with the GW signal $\Delta t$ on the inference of network parameters.}

{To mitigate the impact of $\Delta t$, $t_{0}$ is uniformly distributed in (0.7,0.9) s. Simultaneously, the GW signal is uniformly varied within a range of 0.1 s around, with adjustments ensuring its time remains between 0.7 and 0.9 s. The results are presented in Fig.~\ref{fig10}. Figure~\ref{fig10} illustrates the variation in p-values for parameter inference using the normalizing flow network as $t_{0}$ changes. It depicts error bar for the $1 \sigma$ confidence intervals for 20 sets of results. It shows that with the change in $t_{0}$, the variation of p-values exhibits a relatively smooth trend overall, without significant upward or downward trends. This suggests that a single change in $t_{0}$ does not significantly affect parameter inference robustness.}

{To study $\Delta t$ between the glitch and GW signal, we varied $\Delta t$ in $(-0.1, 0.1)$ s. Glitch times were generated from the GW signal time and $\Delta t$ to assess its impact on parameter robustness inference. The results are shown in Fig.~\ref{fig11}. Figure~\ref{fig11} displays the p-value for network parameter inference as $\Delta t$ varies. The curves in Fig.~\ref{fig11} also represent error bar curves for $20$ sets of data. It can be observed that when $\Delta t$ is relatively large, indicating glitches occurring either before or after the signal, parameter inference remains within a reasonable range. However, when $\Delta t$ is 0 s, indicating overlapping glitches and GW signals, the network parameter inference results are poorer, which is consistent with Ref.~\cite{Macas:2022afm}. This indicates that $\Delta t$ influences network parameter inference, with a significant impact when they overlap.}

{Since a single change in $t_{0}$ does not affect network parameter inference and the overlap between glitches and GW signals has a significant impact, we fixed the glitch time at 0.8 s in this work. This ensures that $\Delta t$ is uniformly distributed in $(-0.1, 0.1)$ s, mitigating its influence on subsequent discussions.}

\bibliography{paper}

\end{document}